\documentclass[runningheads,a4paper]{llncs}
\usepackage{amssymb}
\usepackage{hyperref}
\usepackage{graphicx}
\usepackage{wrapfig}
\usepackage{url}
\usepackage[]{algorithm2e}
\SetKwComment{Comment}{$\triangleright$\ }{}

\newcommand{\keywords}[1]{\par\addvspace\baselineskip
\noindent\keywordname\enspace\ignorespaces#1}

\begin{document}
\mainmatter
\title{EEG-based Subjects Identification based on Biometrics of Imagined Speech using EMD}

\author{Luis Alfredo Moctezuma \and Marta Molinas}
\institute{
Department of Engineering Cybernetics, Norwegian University of Science and Technology. Trondheim, Norway\\
\email{luisalfredomoctezuma@gmail.com, marta.molinas@ntnu.no}
}

\toctitle{EEG-based Subjects Identification based on Biometrics of Imagined Speech}
\tocauthor{Luis Alfredo Moctezuma}
\titlerunning{EEG-based Subjects Identification based on Biometrics of Imagined Speech}
\authorrunning{EEG-based Subjects Identification based on Biometrics of Imagined Speech}

\maketitle
\begin{abstract}
When brain activity is translated into commands for real applications, the potential for human capacities augmentation is promising. In this paper, EMD is used to decompose EEG signals during Imagined Speech in order to use it as a biometric marker for creating a Biometric Recognition System. For each EEG channel, the most relevant Intrinsic Mode Functions (IMFs) are decided based on the Minkowski distance, and for each IMF 4 features are computed: \textit{Instantaneous and Teager energy distribution} and \textit{Higuchi and Petrosian Fractal Dimension}. To test the proposed method, a dataset with 20 subjects who imagined 30 repetitions of 5 words in Spanish, is used. Four classifiers are used for this task - \textit{random forest, SVM, naive Bayes} and \textit{k-NN} - and their performances are compared. The accuracy obtained (up to 0.92 using Linear SVM) after 10-folds cross-validation suggest that the proposed method based on EMD can be valuable for creating EEG-based biometrics of imagined speech for Subjects identification.
\end{abstract}

\section{Introduction}

Electroencephalography (EEG) is a popular non-invasive technique of Brain Computer Interface (BCI), and it refers to the exploration of bioelectrical brain activity registered during different activation functions. EEG does not require any type of surgery, however, compared with invasive techniques the signals obtained are weaker. Another important term is the Electrophysiological source, that refers to the neurological mechanisms adopted by a BCI user to stimulate the brain signals \cite{bashashati2007}.

Due to the easy setup and the little training required, this paper uses the Electrophysiological source \textit{Imagined Speech} \cite{desain2008bci}, that refers to imagined or internal speech without uttering-sounds / articulating-gestures, to create a biometric system  for subjects identification.

Due to the non-stationary and non-linear nature of brain signals, signal processing tools like the Wavelet Transform  \cite{Moctezuma2017a,torres2013analisis,Nishimoto2017} and power spectral density (PSD) \cite{Brigham2010} capable of dealing with these properties, have been reported in the literature. Most recent works have shown wavelets as a powerful tool to analyze brain signals.  However, its main disadvantage is the need to fit the best mother function for the signal. This means that mother functions will be different depending on the task/neuro-paradigm/environment adopted. 

Recently, the Empirical Mode Decomposition (EMD) has been employed to analyze brain signals corresponding to different tasks. It has shown to be robust in decomposing non-stationary and non-linear time series, with the advantage that it does not need a-priory definition of specific parameters to the signal, in contrast with wavelet transform.

The interest in biometric recognition systems has increased in the las years, since traditional security systems (security guards, smart cards, etc) poses serious challenges of increased vulnerabilities. Current biometric security systems are vulnerable due to a variety of attacks to skip the authentication process \cite{Jain2005}. This is because authentication systems cannot discriminate between authorized users and an intruder who fraudulently obtains the access privileges.

To tackle this problem, some researchers have explored the use of brain signals as a measure for a biometric security system. This is possible because any human physiological and/or behavioral characteristic can be used as a biometric feature as long as it satisfies the following requirements: \textit{universality, permanence, collectability, performance, acceptability and circumvention}. A biometric recognition system is able to perform automatic Subjects identification based on their physiological and/or behavioral features\cite{Jain2005} with the advantage that a single biometric trait can be used for the access into several accounts.

In that context, the main neuro-paradigms in the state-of-the-art are: sensorimotor activity, imagination of activities (Visual Counting and geometric figure Rotation \cite{Ashby2011} and mental composition of letters \cite{Palaniappan2006}, among others \cite{Marcos2014}. One of the challenges for subject identification task is the feature extraction stage in order to represent the brain signal captured from different electrodes with a single vector, since it is impractical and  computationally costly to use all data generated by the brain.
%
%
%
%

Some authors report the use of imagined speech, for example \cite{Brigham2010} used EEG signals from a small population of 6 subjects while imagining the syllables \textit{/ba/} and \textit{/ku/}. The collected database consisted of 6 sessions and for each one 20 trials per subject from 128 channels with a sampling frequency of 1024 Hz. using Electrical Geodesics device. For feature extraction they used the PSD for each EEG signal, then autoregressive (AR) model coefficients were computed for each electrode using the Burg method \cite{Steven1988}. The classification stage was performed using the linear kernel of Support Vector Machine (SVM) classifier and using 1-Nearest-Neighbor (k-NN). For these two syllables they obtained 99.76\% and 99.41\% of accuracy respectively. In the work presented in \cite{Nishimoto2017}, resting-states were used for subject identification using \textit{Linear SVM}. The dataset used consisted of 40 subjects, and 192 instances per subject. The sampling frequency was 256 Hz with 64 channels. First, for pre-processing a band-pass filter (0.5-40 Hz) and then the Common Average Reference were applied. For feature extraction the Morlet Wavelet was used to extract power spectrum of 7 frequency bands, to finally apply a downsampling to 32 Hz. The accuracies obtained in the best cases were 100\%, 96\% and 72\% respectively for 3 lengths of the signal (300, 60 and 30 seconds). However, in a real application, the registry of 300, 60 or even 30 seconds of a signal can be impractical and with high computational cost for real-time. In addition the use of 128 or 64 channels does not support the portability of the device.

As a first step to create a robust method without \textit{a-priory} definition of additional parameters, a method based on EMD to extract features from brain signals of Imagined Speech, is presented here.

\subsection{Empirical Mode Decomposition}

The EMD method is useful to decompose non-linear and non-stationary signals into a finite number of Intrinsic Mode Functions (IMFs) that satisfies two conditions \cite{Huang1998}:

 \begin{enumerate}
 \item The number of extrema and the number of zero crossings must be either equal or differ at most by one.
\item At any point, the mean value of the envelope defined by the local maxima and the envelope defined by the local minima is zero.
 \end{enumerate}

The method decomposes a signal into oscillatory components by applying a process called \textit{sifting}. The sifting process for the signal $x(t)$ can be summarized as shown in the algorithm \ref{EMDAlgorithm}:

\begin{algorithm}[H]
{\scriptsize
 \KwData{Time serie = $x(t)$}
 \KwResult{IMFs}
 sifting = True\;
\While{sifting = True}{
 	\begin{enumerate}
 \item Identify all upper extrema in $x(t)$
  \item Interpolate the local maxima to form an upper envelope $u(x)$.
  \item Identify all lower extrema of $x(t)$
  \item Interpolate the local minima to form an lower envelope $l(x)$
  \item Calculate the mean envelope: \\
  		$m(t)=\frac{u(x)+l(x)}{2}$
  \item Extract the mean from the signal:\\
  		$h(t)=x(t)-m(t)$
  \end{enumerate}
  \eIf{$h(t)$ satisfies the two IMF conditions}{
  $h(t)$ is an IMF\;
  sifting = False   \Comment*[r]{Stop sifting}
  }{
  x(t)= h(t)\;
  sifting = True   \Comment*[r]{Keep sifting}
  }
  \eIf{$x(t)$ is not monotonic}{
  \textbf{Continue\;}
  }{
  \textbf{Break\;}
  }
  }
  }
  \caption{The sifting process for a signal $x(t)$}\label{EMDAlgorithm}
\end{algorithm}

\subsection{IMFs selection}

The EMD is a powerful tool to decompose a non-stationary signal, however some IMFs that contain limited information may appear in the decomposition because the numerical procedure is susceptible to errors \cite{Rilling2003}. To select the IMFs that contain the most relevant information about the signal, the methods presented in \cite{Souza2014,Boutana2010} were applied and compared, to finally use the method proposed in \cite{Boutana2010} that employs the Minkowski Distance ($d_{mink})$, as follow.

	\begin{equation}
		d_{mink} =\Biggl(\sum_{i=1}^{n} \big|x_{i}-y_{i}\big|^2  \Biggr)^{1/2}
	\end{equation}

where $x_i$ and $y_i$ are the $i$-th respective samples of the observed signal and the extracted IMF.

According to the authors, the redundant IMFs have a shape and frequency content different than those of the original signal, which means that when a IMF is not appropriate, the $d_{mink}$ presents a maximum value.
\medskip

In this work, a new method for feature extraction based on EMD is proposed. In the next section the method is described in brief. Then, the application of the proposed method for Subjects identification is explained.

\section{Description of the method}

The main contribution of the proposed method is the feature extraction stage, that consists on applying the Empirical Mode Decomposition (EMD) to obtain 5 Intrinsic Mode Functions (IMF) per channel of the EEG data. 

To select the most relevant IMFs, the Minkowski Distance was computed \cite{Boutana2010}. Once the most relevant IMFs from all instances in the dataset were obtained, it turned out that the number of IMFs was different depending of the size of the signal and the imagined word. However, to obtain meaningful features it is necessary to have the same number of IMFs from all instances. To cope with this, the IMFs selected were limited to the minimum relevant IMFs in all instances, which in this case were only the first 2 IMFs.

\begin{figure}
\centering
\includegraphics[scale=0.435]{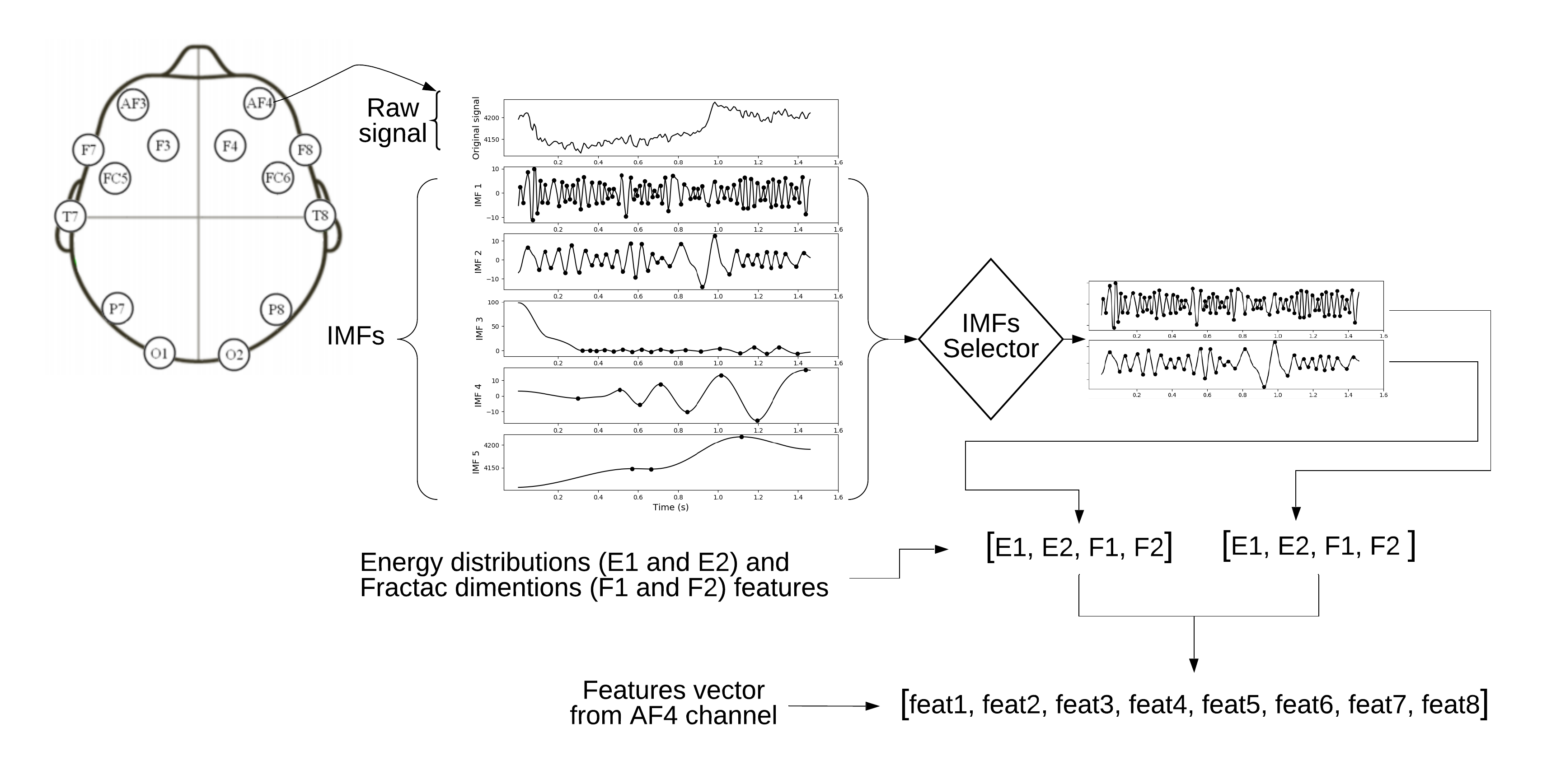}
\caption{Flowchart summarizing the feature extraction stage.}
\label{fig:Flowchart}
\end{figure}

Then for each IMF, 4 features were computed: \textit{Instantaneous energy, Teager energy, Higuchi fractal dimension } and \textit{Petrosian fractal dimension}, as it is shown in the figure \ref{fig:Flowchart}. All features per IMF and per channel were concatenated to obtain a feature vector per instance. Once the feature vectors were obtained, they were used to train 4 machine learning-based classifiers (\textit{random forest, naive Bayes, Support Vector Machine (SVM)} and \textit{K-Nearest Neighbors (k-NN)}) in order to compare their performances.

In the following, the step-by-step procedure proposed in this work to identify Subjects by using the EEG of their imagined speech, is described.

\subsection{Feature extraction}
When the most relevant IMFs were selected, 4 features were computed for each one in order to reduce the feature vector to obtain a good representation of the signal. The first 2 features used are related with the energy distribution and the others two with Fractal dimensions. Each feature tested in this work is here described in brief.

\begin{itemize}
\item \textsc{Instantaneous Energy}: gives the energy distribution in each band \cite{EDidiot}:
			\begin{equation}
				 f_{j} =log_{10}\Biggl( \frac{1}{N_{j}} \sum_{r=1}^{N_{j}} (w_{j}(r))^2 \Biggr) 
			\end{equation}

\item \textsc{Teager Energy}: This energy operator reflects variations in both amplitude and frequency of the signal and it is a robust parameter as it attenuates auditory noise \cite{EDidiot,Jabloun1999}.
	\begin{equation}	
		f_{j} =log_{10}\Biggl( \frac{1}{N_{j}} \sum_{r=1}^{N_{j}-1} \Big|(w_{j}(r))^2 -w_{j}(r-1)*w_{j}(r+1) \Big|  \Biggr)
	\end{equation}

\item \textsc{Higuchi Fractal Dimension}: The algorithm approximates the mean length of the curve using segments of \textit{k} samples and estimates the dimension of a time-varying signal directly in the time domain \cite{Higuhi1988}. Considered a finite set of observations taken at a regular interval: $X(1), X(2), X(3), .., X(N)$. From this series, a new one $X_{k}^{m}$ must be constructed,
	\begin{equation}
		X_{k}^{m}: X(m), X(m+k), X(m+2k), .., X\biggl(m+ \biggl(\frac{N-m}{k}\biggl) k\biggl)
	\end{equation}
    Where $m = 1, 2,.., k$, $m$ indicate the initial time  and $k$ the interval time.
    Then, the length of the curve associated to each time series $X_{k}^{m}$ can be computed as follow:
    	\begin{equation}	
		L_{m}(k) = \frac{1}{k}\Bigg( \sum_{i=1}^{\frac{N-m}{k}} \bigg(X(m+ik)-X\Big(m+ (i-1)k\Big)\bigg)  \Bigg) \Bigg(  \frac{N-1}{\Big(\frac{N-m}{k} \Big)k }\Bigg)
	\end{equation}
    Higuchi takes the mean length of the curve for each $k$, as the average value of $L_{m}(k)$, for $m=1, 2, ..., k$ and $k=1, 2, ..., k_{max}$, that it is calculated as:
	\begin{equation}
		L(k) = \frac{1}{k} \sum_{m-1}^{k} (L_{m}(k))
	\end{equation}

\item \textsc{Petrosian Fractal Dimension}: can be used to provide a fast computation of the fractal dimension of a signal by translating the series into a binary sequence \cite{Petrosian1995}.
	\begin{equation}
		FD_{Petrosian} = \frac{\log_{10}n }{\log_{10}n+\log_{10} \bigg(\frac{n}{n+0.4N_{\nabla}} \bigg)}
	\end{equation}
    Where $n$ is the length of the sequence and $N_{\nabla}$ is the number of sign changes in the binary sequence.
\end{itemize}

\subsection{Classifiers and validation}

At this point, the features vector have the same features per each instance with an assigned tag. This allows the use of machine learning methods. In this work,  the machine learning methods \textit{random forest, naive Bayes, SVM} and \textit{k-NN} were used. For \textit{SVM}, all experiments were reproduced with the kernels \textit{Linear, RBF (Radial Basis Function)} and \textit{Sigmoid}. In the \textit{random forest} case, the experiments were reproduced with different tree depths (2, 3, 4, 5) using the \textit{Gini} impurity. \textit{k-NN}  was tested with different number of neighbors (1, 2, 3, 4, 5, 6, 7, 8, 9).

The accuracy with the 4 classifiers was estimated to evaluate their performances using 10-folds cross-validation.

%
%

\section{Dataset and experiments}

%
%

In this section, the dataset used to test the proposed method in two different experiments for Subjects identification is described in brief . 

The purpose of the first experiment is to show that the Subject can be identified regardless of the imagined word, to show if a biometric system using different password per Subject can be possible. The second experiment aim is to find whether the accuracy is higher if the password is pre-defined. In others words, using the same imagined word as biometric security measure.

%
%


%
%

\subsection{Dataset}

The dataset consists of EEG signals from 27 subjects recorded using EMOTIV EPOC device while imagining 33 repetitions of five imagined words in Spanish; \textit{arriba, abajo, izquierda, derecha} and \textit{seleccion}, corresponding to the English words \textit{up, down, left, right} and \textit{select}.
%
%
%
%

Each repetition of the imagined words was separated by a state of rest. The protocol for EEG signal acquisition is described in details in \cite{torres2013analisis}.

EEG signals were recorded from 14 channels which were placed on the head according to the 10-20 international system \cite{Jasper1958}, with a sample frequency of 128 Hz.

\subsection{Setup}

%
%

For the next experiments the first 20 subjects and the first 30 repetitions per each of the 5 imagined words were used . In summary, the terms used along the paper are the following.
%
%

\begin{itemize}
\item $S_{\nabla}=20$: Subjects.
\item $W_{\nabla}=5$: The imagined words in the dataset.
\item $R_{\nabla}=30$: Repetitions per imagined word.
\item $C_{\nabla}=14$: Channels used in all instances.
\item $IMFs_{\nabla}=2$: IMF per channel.
\item $F_{\nabla}=4$: Features per IMF,  corresponding to \textit{Teager energy, Instantaneous energy, Higuchi Fractal Dimension} and \textit{Petrosian Fractal Dimension}.
\end{itemize}

According to the proposed method the feature vector size per subject is $F_{\nabla}*IMFs_{\nabla}*C_{\nabla} = 112$. Next, the specific setup for the experiments and the results are presented.

\subsection{Subject level analysis}

In this experiment, 4 classifiers were used in order to compare their performances and each classifier has $S_{\nabla}$ classes, and per class $R_{\nabla}*W_{\nabla}=150$ instances.

In the table \ref{table:allWordsT} the accuracies obtained with the proposed method are shown. 

\begin{table}[!ht]
\begin{center}
\caption{Accuracy obtained when all imagined words were joined for subjects identification}
\setlength\tabcolsep{3pt}
\begin{tabular}{lc}
\hline
Classifier& Accuracy\\
\noalign{\smallskip}
\hline
\noalign{\smallskip}
random forest&0.64\\
SVM&0.84\\
naive Bayes&0.68\\
k-NN& 0.78\\ \hline
\end{tabular}
\label{table:allWordsT}
\end{center}
\end{table}

The aim of this experiment is to show that the method can be used for subjects identification with high accuracy rates. According to the results in the table \ref{table:allWordsT}, the classifier SVM was the best. As it was mentioned before, SVM was tested with different kernels and for this experiment the best one was the \textit{Linear SVM}.

\subsection{Word level analysis}

In this experiment, the classifiers were trained using all words separately, in order to explore if the proposed method works best for a specific word. Each classifier has $S_{\nabla}$ classes, and per class $R_{\nabla}$ instances.

\begin{table}[!ht]
\begin{center}
\caption{Accuracy obtained per imagined word for subjects identification}
\setlength\tabcolsep{3pt}
\begin{tabular}{lccccc}
\hline
Classifier& Up& Down& Left& Right& Select\\
\noalign{\smallskip}
\hline
\noalign{\smallskip}
random forest&0.78&0.77&0.73&0.73&0.75\\
SVM&0.91&0.87&0.88&0.84&0.92\\
naive Bayes&0.90&0.85&0.88&0.85&0.89\\
k-NN&0.85&0.80&0.81&0.79&0.88\\ \hline
\end{tabular}
\label{table:allWords}
\end{center}
\end{table}

In this experiment also the highest accuracy was obtained when using the \textit{SVM} classifier with the \textit{Linear} kernel, obtaining the accuracy of 0.92. On the other hand, the lowest performance was obtained using the classifier \textit{random forest}

\section{Discussion and Conclusions}

In this paper, a method based on EMD for Subjects identification from EEG signals of imagined speech was presented. The proposed method was applied to a dataset of Imagine Speech with encouraging results. The accuracies obtained suggest that the the use of imagined speech for Subjects identification, specially using the classifier \textit{Linear SVM}, can be effective and worth exploring further. 

When the imagined words were joined to observe if it is possible to identify a Subject regardless of the imagined word, the highest accuracy obtained was 0.84 using \textit{Linear SVM}. Then, when the second experiment was carried out to observe the accuracies using the imagined words separately, the maximum accuracies  reached were also using \textit{Linear SVM}: 0.91, 0.87, 0.88, 0.84 and 0.92 respectively per each imagined word.

%
%
%
%

In the work presented in \cite{Moctezuma2018} the Common Average Reference (CAR)\cite{Bertrand1985} was used to improve the signal-to-noise ratio. Then, the feature extraction was based on \textit{Instantaneous} and \textit{Teager} energy distribution of 4 decomposition levels of Wavelet using the mother function  Biorthogonal 2.2, and \textit{random forest} for classification.  The accuracies obtained using the imagined word \textit{select} were 0.96 and 0.93 respectively. In this paper, the highest accuracy (Using Linear SVM) obtained for the imagined word \textit{select} was 0.92,  which is slightly lower than the above ones. However, with the inherent adaptivity of EMD for feature extraction, there is no need to pre-define any mother function for the particular task or neuro-paradigm. In addition, the EMD inherently improves the signal-to-noise ratio by removing the noise in the first IMFs.

A limitation of the proposed method is the use of a dataset of brain signals from only 20 subjects. In future, it will be necessary to reproduce the experiments using a larger population in order produce an alternative competitive system to the current biometric security systems used in industry. Further efforts will be made to explore alternative  techniques for IMFs selection and for EEG channels selection, since it is well known that specific channels will provide more relevant information than others for the distinct task selected for subject identification.
\subsubsection*{Acknowledgments.}
This work was supported by Enabling Technologies - NTNU, under the project ``David versus Goliath: single-channel EEG unravels its power through adaptive signal analysis - FlexEEG''.


\end{document}